%% file: main-authordraft.tex
\documentclass[sigconf]{acmart}
\usepackage{graphicx} 
\usepackage{multirow}

\AtBeginDocument{%
  \providecommand\BibTeX{{%
    \normalfont B\kern-0.5em{\scshape i\kern-0.25em b}\kern-0.8em\TeX}}}

\copyrightyear{2025}
\setcopyright{cc}
\setcctype{by-nc}
\acmConference[Preprint] {Summer, 2025}

\begin{document}

\title{African Data Ethics: A Discursive Framework for Black Decolonial Data Science}

\author{Teanna Barrett}
\affiliation{%
  \institution{Paul G. Allen School of Computer Science \& Engineering}
  \institution{University of Washington}
  \city{Seattle}
  \state{Washington}
  \country{USA}
}
\email{tb23@cs.washington.edu}

\author{Chinasa T. Okolo}
\affiliation{%
  \institution{The Brookings Institution}
  \city{Washington}
  \state{D.C.}
  \country{USA}}
\email{cokolo@brookings.edu}

\author{B. Biira}
\affiliation{%
  \institution{Information School}
  \institution{University of Washington}
  \city{Seattle}
  \state{Washington}
  \country{USA}
}
\email{bbiira@uw.edu}

\author{Eman Sherif}
\affiliation{%
 \institution{Paul G. Allen School of Computer Science \& Engineering}
 \institution{University of Washington}
  \city{Seattle}
  \state{Washington}
  \country{USA}}
\email{emans@uw.edu}

\author{Amy X. Zhang}
\affiliation{%
  \institution{Paul G. Allen School of Computer Science \& Engineering}
  \institution{University of Washington}
  \city{Seattle}
  \state{Washington}
  \country{USA}}
\email{axz@cs.uw.edu}

\author{Leilani Battle}
\affiliation{%
  \institution{Paul G. Allen School of Computer Science \& Engineering}
  \institution{University of Washington}
  \city{Seattle}
  \state{Washington}
  \country{USA}}
\email{leibatt@cs.washington.edu}

\renewcommand{\shortauthors}{Barrett et al.}

\input{sections/00_abstract}



\maketitle
\input{sections/01_introduction}
\input{sections/02_background}
\input{sections/03_methods}
\input{sections/04_results}
\input{sections/05_discussion}
\input{sections/06_conclusion}


\begin{acks}
Thank you to Joseph Asike, George Obaido, Innocent Obi Jr., and Andrew Shaw for additional advice and support. The work presented was funded in part by the NSF (award \# 2141506), the Sloan Foundation, and Google.
\end{acks}
\bibliographystyle{ACM-Reference-Format}
\bibliography{main-authordraft}

\appendix
\input{sections/07_appendix}

\end{document}

%% file: sections/00_abstract.tex
\begin{abstract}
 The shift towards pluralism in global data ethics acknowledges the importance of including perspectives from the Global Majority to develop responsible data science practices that mitigate systemic harms in the current data science ecosystem. Sub-Saharan African (SSA) practitioners, in particular, are disseminating progressive data ethics principles and best practices for identifying and navigating anti-blackness and data colonialism. To center SSA voices in the global data ethics discourse, we present a framework for African data ethics informed by the thematic analysis of an interdisciplinary corpus of 50 documents. Our framework features six major principles: 1) Challenge Power Asymmetries, 2) Assert Data Self-Determination, 3) Invest in Local Data Institutions \& Infrastructures, 4) Utilize Communalist Practices, 5) Center Communities on the Margins, and 6) Uphold Common Good. We compare our framework to seven particularist data ethics frameworks to find similar conceptual coverage but diverging interpretations of shared values. Finally, we discuss how African data ethics demonstrates the operational value of data ethics frameworks. Our framework highlights Sub-Saharan Africa as a pivotal site of responsible data science by promoting the practice of communalism, self-determination, and cultural preservation.
\end{abstract}

%% file: sections/01_introduction.tex
\section{Introduction}
Artificial intelligence (AI), machine learning (ML), and data science research and development are primarily conducted in economically powerful countries such as the United States, China, the United Kingdom, France, and Germany \cite{bengio2024international}. 
Consequently, data-driven technology (DDT) leveraging these methodologies often reflect the logics of large corporations and the hegemonic values of the countries they are based in \cite{birhane2022forgotten}. 
When ethical dilemmas arise within data science development, mitigation strategies are often limited by the perspectives of these actors who rarely experience the direct consequences of data harm \cite{mhlambi2023decolonizing,cisse2018look,gwagwa2022role}.

To balance these dominant perspectives, pluralistic data ethics calls for perspectives from the Global Majority to 
better mitigate systemic concerns \cite{rifat2023many,carman2023applying}. 
Embracing pluralism in global data ethics acknowledges the limits of any single ethical perspective in guiding comprehensive responsible data science (RDS) practices. 
Unfortunately, Sub-Saharan African (SSA) \footnote{Throughout the rest of the document, Africa or SSA will be usedACM FAcct 2025 Proceedings to refer to Sub-Saharan Africa.} voices have routinely been left out of data science ethics discussions \cite{eke2023responsible}. 
This is a glaring omission---African data collaborators have an intimate experience with how anti-blackness and data colonialism operate in data science work \cite{abebe2021narratives,eke2023responsible,brevini2024critiques}.
African people are often exploited as data workers, their resources are extracted for computing infrastructure, and even when deploying systems in their own communities, many processes are managed and owned by external actors \cite{birhane2020algorithmic}.

To address the underrepresentation of African perspectives in global data ethics, African data scientists have returned to African philosophies such as Ubuntu to articulate the current ethical dilemmas in data science work on the continent \cite{gwagwa2022role}. 
However, African data ethics texts are spread across a variety of publication venues and are rarely synthesized into a cohesive review or framework \cite{eke2022responsible}. For this reason, it can be difficult to grasp common topics, differing fields of thought, and how proposed principles translate to current practices of RDS in Africa. In this work, we synthesize the current African data ethics discourse and contextualize data ethics theories through the application of African philosophies. We conduct a thematic analysis of 50 documents from African philosophy, information ethics, AI ethics, and human-computer interaction (HCI) research to answer the following research question: \textbf{What are the African values and ethical theories that can inform responsible data science (RDS) practices in Africa?} 
As a result of our inquiry, we derived an ethical framework consisting of six major principles rooted in the realities of data science work in Africa (Table~\ref{tab:framework-overview}). 

Our framework surfaces novel understandings, theories, and practices for RDS in Africa that contend with the legacy of colonialism and the current administration of neocolonialism to articulate the inefficacy of current data science relations in African contexts. 
Several principles also engage with a range of African communitarian theories to recontextualize common paradigms of social good, participatory design, and comprehensive data ethics education. The framework also exhibits the practical solutions African data scientists offer to local governments, multinational technology companies, grassroots collectives, and individual data scientists engaged in the African data science ecosystem. 

We place our framework in conversation with other particularist data ethics frameworks to gauge its unique contributions to global efforts. While every framework is concerned with implementing technology for social good, we find African data ethics problematizes vague universalist language and calls for material redress for data harms.
Overall, our work aims to bridge rich African philosophical traditions with contemporary data science practices, developing a framework that can guide responsible technology development within African contexts while contributing to global discussions on data ethics and justice.

%% file: sections/02_background.tex
\section{Background}
\label{sec:background}

\subsection{Pluralistic Efforts in Global Data Ethics}
Data ethics encompasses the normative frameworks and moral principles governing the collection, processing, storage, and deployment of data \cite{floridi2016data}. 
An ethical principle, in a data ethics context, represents a \textbf{fundamental normative guide for action}---such as respect for persons, beneficence, or justice---that transcends mere technical guidelines or professional best practices. What distinguishes genuine ethical principles from operational procedures are their foundation in moral philosophy and their universal applicability across contexts.
Literature within this field has proposed new data ethics paradigms \cite{krijger2023ai, verhulst2021reimagining, burr2023ethical}, critiqued current approaches to responsible data science \cite{taylor2019responsible, leonelli2016locating, gansky2022counterfacctual}, and experimented with algorithmic and other technical approaches to mitigate bias \cite{hort2024bias, wang2023mitigating, smith2020mitigating}. While scholarly communities such as the ACM Conference oACM FAcct 2025 Proceedingsn Fairness, Accountability, and Transparency (FAccT) and the AAAI Artificial Intelligence, Ethics, and Society (AIES) serve as platforms for engaging discourse on AI and data ethics, a significant portion of the scholarship produced in these venues is primarily rooted in Western philosophy. 

While Western philosophical traditions have dominated data ethics discourse through frameworks such as utilitarianism and deontological ethics \cite{moran2022towards, heavin2024digital}, emerging scholarship increasingly recognizes the necessity of incorporating diverse ethical paradigms and epistemological frameworks from Global Majority perspectives. For example, Indigenous knowledge systems provide alternative conceptualizations of data sovereignty and stewardship that challenge Western individualistic notions of privacy and ownership \cite{carroll2023care, hudson2023indigenous, walter2021indigenous, lovett2019good, rainie2019indigenous}. East Asian philosophical traditions, including Confucian ethics, contribute valuable insights regarding social harmony and collective responsibility in technological development, offering nuanced frameworks for balancing individual rights with communal interests in data-driven systems \cite{d2023ai, ess2006ethical, yeh2010effect}. Traditional African philosophy offers Ubuntu-based approaches that emphasize collective well-being and communal responsibility in data ethics \cite{reviglio2020datafied, okyere2023place}. HCI and data ethics scholars have also applied knowledge from Black studies such as Black feminism \cite{reece2025powerful}, critical race theory \cite{coleman2024blk}, Afropessimism \cite{cunningham2023grounds}, prison abolitionism \cite{benjamin2019race}, and historical and contemporary social analysis \cite{gebru2024tescreal, noble2018algorithms, browne2015dark, monroe-white2021emancipatory} to articulate how data technology perpetuates anti-black racism in the United States.

\subsection{African Philosophy: Foundation of African Data Ethics}
 \textbf{African Philosophy is a sub-domain of philosophy meant to reclaim and generate philosophical theories from Africans \cite{ramose2004struggle}.} Around the time of widespread decolonization, African philosophers pivoted away from classic Western philosophy to uncover the rich intellectual theories and practices of pre-colonial Africans \cite{wiredu2004akan}. Beyond examining the past, African philosophers have also worked to cultivate an intellectual home for Africans to generate new philosophies that speak to the realities of African life in modernity \cite{ramose2004struggle, kohnert2022machine}. 

Several African philosophical frameworks offer valuable insights for data ethics and responsible technology development. The Yoruba concept of ``iwa'' (character/moral behavior) emphasizes the ethical implications of one's actions on the collective community, suggesting approaches to data governance that prioritize communal benefit over individual gain \cite{oyeshile2021yoruba}. Similarly, the Akan concept of ``onipa'' (personhood) \cite{wingo2006akan, owusu2019onipa} and the Zulu notion of Ubuntu, often translated as ``umuntu ngumuntu ngabantu'' (a person is a person through other persons) \cite{reviglio2020datafied, gwagwa2022role}, provide frameworks for understanding human dignity and agency in technological contexts, particularly relevant for issues of consent and data sovereignty. The Ethiopian philosophy of ``Medemer'' (synergy/coming together) offers a model for collaborative data sharing and governance that balances individual autonomy with collective benefit \cite{assefa2024critical}. Additionally, the Igbo concept of ``omenala'' (customs/traditions) \cite{nwoye2011igbo, nwala1985igbo} and the Swahili principle of ``ujamaa'' (familyhood) \cite{nyerere1962ujamaa} suggest that data governance should align with existing social structures and cultural practices rather than imposing external frameworks. Overall, these philosophies collectively lay the foundation for ethical frameworks that challenge Western individualistic approaches to data privacy, data sharing, and ownership while promoting interdependence and cultural alignment.

\subsection{Current AI Ethics Discourse in Africa}
A growing number of technology researchers are leveraging African philosophy to assess technological development and AI \cite{ewuoso2021african, capurro2008information, ruttkampbloem2023epistemic, gwagwa2022role, attoe2023conversations}. 
This work, along with foundational African philosophy, can inform practical approaches to data collection protocols that respect communal ownership and decision-making \cite{buhler2023unlocking}, inspire algorithmic fairness metrics that incorporate African conceptions of justice and equity \cite{asiedu2024case}, and help shape privacy policies that balance individual rights with community interests \cite{jimoh2023quest}. Additionally, African philosophy can be valuable in informing ethics review processes that consider local cultural contexts and values, and help build data governance structures that reflect African leadership models and decision-making practices \cite{chinweuba2019philosophy}. Our work aims to help bridge these issues and expand discourse on how African philosophies can contribute to advancing responsible data science practices within the continent.

\subsubsection{Practitioner-led}
Sábëlo Mhlambi's pioneering work on African data ethics, grounded in the principles of Ubuntu \cite{mhlambi2020from}, draws from the epistemic, ontological, and ethical theories of African philosophers like Mogobe Ramose \cite{ramose2004ethicsofubuntu} to critique prevailing AI paradigms \cite{mhlambi2023decolonizing}. 
Since Mhlambi's Ubuntu data ethics contribution, a significant amount of African scholarship on data ethics has highlighted Ubuntu as an African philosophy that should be engaged with more in global data ethics work \cite{gwagwa2022role, segun2021critically, langat2020how, kiemde2022towards, goffi2023teaching}. 
This body of work makes a general appeal to African communitarian ethics to critique current ethical paradigms \cite{metz2021african}, report on data science work in Africa \cite{day2023data}, and survey perceptions and concerns among African data science practitioners \cite{eke2022forgotten}. There is also a plethora of African data ethics gray literature published through African data organizations (often in partnership or funded by institutions in Europe or the United States) by way of blog posts \cite{shilongo2023creativity}, reports \cite{sinha2023principlesafrofeminist}, and formalized briefs \cite{gwagwa2019recommendations}. 
It must also be noted that many contributions made by non-African data ethicists disproportionately fail to engage with other African philosophies beyond Ubuntu. 

\subsubsection{Continental \& Local Policy}
The development of local and continental policies for AI and data regulation can help African countries improve adherence to data ethics frameworks while steering responsible AI development. 
Currently, 38/55 African Union (AU) member states have enacted formal data protection regulations, with Malawi and Ethiopia recently enacting data protection laws in mid-2024 \cite{okolo2024operationalizing}. The African Union has also released continental frameworks, such as the African Union Convention on Cyber Security and Personal Data Protection (Malabo Convention), which mandates national cybersecurity policies and strategies while addressing personal data protection and cybercrime \cite{malaboconvention}. 
In parallel, efforts to regulate AI are gaining momentum, with 14 countries adopting national AI strategies \cite{ECDPM-Africa}, complemented by the AU’s AU-AI Continental Strategy published in August 2024 \cite{african_union2024continental}. However, systematic gaps in data regulation persist \cite{okolo2024operationalizing, eke2022responsible, john2021technology, osakwe2021strengthening}, and if left unaddressed, these shortcomings may undermine AI regulatory efforts. To advance data ethics and safeguard African communities, governments must prioritize enforcement of existing data privacy laws while ensuring regulations provide adequate protections.

%% file: sections/03_methods.tex
\section{Methods}
\label{sec:methods}

Inspired by ongoing work, we have two aims in developing an ethical framework for African data ethics: 1) honoring the existing scholarship of African data ethicists by preserving distinct perspectives in the discourse, and 2) re-engaging with classic African philosophies to productively expand on African data ethics principles. To achieve the balance between a respectfully discursive and productively expansive framework, we combined methods from qualitative document analysis and literature reviews to develop our qualitative analysis protocol \cite{morgan2022conducting, birhane2022forgotten, battle2024what}. 

\subsection{Data Collection}
The first author seeded the search by reading \textit{The African Philosophy Reader} \cite{coetzee2004african} due to prior exposure.
This text informed subsequent keyword searches in established academic databases: Google Scholar, Web of Science, Scispace,\footnote{
Scispace is a language model search engine for literature reviews: \url{https://typeset.io}} and the publication repositories of ACM and IEEE.
The first author also searched their institutional library and online African philosophy libraries accessible through their institution.
They used key phrases such as \textbf{``African AI ethics''}, \textbf{``African philosophy''} and \textbf{``African data ethics''} to identify relevant literature. 
Documents from database searches were excluded if African values and data science practices were not the primary topic. 
In addition, documents were excluded if they were not a full document. Full documents were understood as non-archival and archival papers (no extended abstracts), reports, or book chapters. No range was set on the publishing year to permit the inclusion of foundational texts from African philosophy.

In parallel to keyword searches, we requested literature recommendations from other scholars in the field. In addition, we used reference and citation tracking to identify relevant documents missed in searches. By the end of our iterative data collection, 50 documents were collected. The details of document inclusions and exclusions can be found in Appendix~\ref{apdx:method}.

\subsection{Thematic Analysis}
 The first author reflexively coded the documents through a practice iteration followed by two rounds of coding to surface themes of African data ethics. 
First, the first author began a grounded reflection process by selecting six documents as a representative set of corpus topics \cite{coetzee2004particularity, mhlambi2020from, african_union2024continental, ndjungu2020blood, segun2021critically, day2023data}.
From this selection, she highlighted and recorded meaningful excerpts from each document. Then she noted how the excerpt answers the research question. These reflections enabled the first author to focus on coding implicit or explicit African ethical principles. In subsequent coding iterations, we consider an African ethical principle to be a \textbf{moral value, understanding, or standard prioritized by or derived from SSA communities}. 
The first author then repeated the above process for each document in the full corpus, in which she coded excerpts that discussed African ethical principles relevant to RDS. Then, she organized the resulting codes into themes through affinity diagramming \cite{scupin2008kj} using Miro.\footnote{\url{https://miro.com/}} 
After the first full round of coding and principle clustering, she identified several minor principles with conceptual gaps that required the analysis of additional documents. 

\subsection{Positionality Statement}
Our research team's converging individual interests have led us to develop and propose our framework for African data ethics. The second author has extensive experience in developing AI policy in Africa and research expertise in developing explainable AI paradigms for marginalized communities. Other authors identify with a variety of African diasporic communities and are dedicated to expanding the representation of Black data science communities in their work on climate change, computer science education and responsible data science. For the author who does not belong to the African Diaspora, their work in pluralistic alignment of AI necessitates learning about the values of all user communities especially those on the margins. Since our research team is affiliated with academic and independent institutions based in the United States and the majority are not directly involved in technology collectives or disciplines represented in our corpus, the lens of epistemic humility guided our framework construction.

The first author is a Black, Jamaican-American woman. She acknowledges that her recent ethnicity is not Sub-Saharan African, but is committed to engaging and contributing to African data ethics. Her positionality as United States-educated, English monolinguist and observer of African data ethics discourse limits generative efforts. Her analytical perspective may also perpetuate misconceptions of the myriad of African perspectives highlighted in the work. As such, she aimed to reflexively guide the study by seeking guidance from technologists, policymakers and thought leaders within African data science communities. 

%% file: sections/04_results.tex
\section{African Data Science Ethical Framework}
In this section, we summarize the key components of our framework for African data ethics. For an overview of the six major principles, 19 associated minor principles and recommended practices or policies derived from our analysis, view Table ~\ref{tab:framework-overview}.

\begin{table*}
    \centering
    \resizebox{\textwidth}{!}{
    \begin{tabular}{| p{0.15\textwidth} | p{0.3\textwidth} | p{0.55\textwidth} |}
        \hline
        Major Principle & Minor Principle & Practices \& Policies\\
        \hline
        \multirow{2}{*}{\parbox{0.15em}}{Challenge Power Asymmetries} & Challenge Colonial Power & Adopt data science practices that are localized, decentralized, bottom-up and in less deference to Big Tech paradigms \\
        \cline{2-3}
        & Challenge Internal Power Asymmetry & Establish agencies to triage citizen concerns and doubts about new technology \\
        \hline
        \multirow{3}{*}{\parbox{0.15em}}{Assert Data Self-Determination} & "For Africans, By Africans" & Africans speak for themselves in global data ethics discourses, negotiations for technical partnerships and design for local needs\\
        \cline{2-3}
        & Treasure Indigenous Knowledge & Conduct natural language processing projects to document and represent indigenous languages \\
        \cline{2-3}
        & Protect Data Agency & Integrate communal conceptions of privacy for data protection such as data trusts, data collaboration tools, and community data management\\
        \hline
        \multirow{3}{*}{\parbox{0.15em}}{Invest in Local Data Institutions \& Infrastructures} & Invest in Physical \& Organizational Infrastructures & Construct secure and locally-run peer to peer networks\\
        \cline{2-3}
        & Cultivate Governance Infrastructure & Develop protocol for conducting algorithmic impact assessments\\
        \cline{2-3}
        & Support Formal \& Informal Data Science Collectives & Provide technical training and education for data scientists outside of the university and working in the informal sector\\
        \hline
        \multirow{3}{*}{\parbox{0.15em}}{Utilize Communalist Practices} & Ground Design in Community-Engaged Consensus & Consult intended users on a routine basis to build trust and incorporate their cultural perspectives in design\\
        \cline{2-3}
        & Instantiate Reciprocal Relationships & Collaborate with African organizations already conducting research or building technologies of interest\\
        \cline{2-3}
        & Mend Data-Driven Harms & Implement reconciliation processes that place disenfranchised communities in positions of power and hold the perpetrating parties accountable\\
        \hline
        \multirow{4}{*}{\parbox{0.15em}}{Center Communities on the Margins} & Design in Solidarity  & Embrace diverse data ethics standards and avoid imposing standards on other communities\\
        \cline{2-3}
        & Center Remote \& Rural Communities & Develop AI solutions that address the needs of remote communities such as agriculture, climate change, and healthcare accessibility\\
        \cline{2-3}
        & Center Women & Incentivize women-led data science entrepreneurship\\
        \cline{2-3}
        & Protect \& Empower Youth & Organize continental meetings for young Africans to discuss the state of data science in their communities\\
        \hline
        \multirow{4}{*}{\parbox{0.15em}}{Uphold Common Good} & Adopt Technology with a Measured Mindset & Appraise imported state-of-art technologies before use \\
        \cline{2-3}
        & Preserve the Dignity of Data Contributors & Treat every collaborator as morally responsible agents deserving of materially and psychologically safe working conditions\\
        \cline{2-3}
        & Strive for Common Good with Systemic Change & Build open data ecosystems that safeguard against harmful use and compensate data subjects from economic gains\\
        \cline{2-3}
        & Maintain Harmony with Stakeholders & Update country-level data ethics to align with continental frameworks and vice versa\\
        \hline
    \end{tabular}
    }
    \caption{Overview of the major and minor principles of our proposed African data ethics framework as well as highlighted practices and policies associated with each minor principle.}
    \Description{A table of the major, minor and practices/policies surfaced from our African Data Ethics corpus. The ordering of major and minor principles follows the order in section 4.}
    \label{tab:framework-overview}
\end{table*}

\subsection{Challenge Power Asymmetries}
Challenging power structures in technological development is not only necessary to mitigate the perpetuation of colonial power legacies, but also misuse and exploitation by any authority. 

\textbf{Challenge Colonial Power.}
\label{sec:chall_colo}
RDS practices from the West do not seamlessly transfer to the African context
because these practices are developed within colonial contexts disconnected from the realities of African practitioners and users \cite{eke2022forgotten, shilongo2023creativity,adelani2022masakhaner,gwagwa2019recommendations,eke2023introducing,okolo2023responsible, eke2023towards, goffi2023teaching,carman2023applying}. African practitioners identify three dimensions in which the sociotechnical disconnect harms responsible data science in Africa: epistemic injustice, dehumanizing extraction, and dependent partnerships. 
Firstly, African scholars identify how philosophical epistemic injustice permeates global data ethics paradigms \cite{eke2022forgotten, metz2021african, olojede2023towards}. As many African philosophers agree, Enlightenment ideals (a premier part of the Western philosophical canon) were predicated on colonialism and anti-black racism \cite{gwagwa2019recommendations}. European colonizers weaponized the Enlightenment emphasis on human rationality to deem Africans sub-human by assessing them as primitive and incapable of reason \cite{lauer2017african}. Colonization was not only justified but encouraged by the philosophies of rationality. Colonizers determined they had a moral obligation to develop Africans capacity for reason through European education. The legacy of colonialism is why African data scientists encourage casting aside Western perspectives to develop distinctly African RDS perspectives \cite{mhlambi2020from}. Additionally, an over reliance on ``rational'' performance metrics encourages the same epistemic injustice that justifies and encourages the marginalization of Africans in technology such as facial recognition \cite{mhlambi2020from, buolamwini2018gender, cisse2018look, gwagwa2022role}. 

Secondly, many documents recognize most African contributions to data science as disproportionately benefiting corporations such as OpenAI, Google, Meta, and Microsoft \cite{ndjungu2020blood,chan2021limits,abebe2021narratives, nwankwo2019africa, kiemde2022towards}. The computing demand of large-data systems such as AI proliferates neocolonialism to new heights in Africa \cite{eke2023towards}. The work of Africans within the data science ecosystem should benefit Africans first \cite{kohnert2022machine}. The fact that it currently does not is connected to the legacy of colonialism and chattel slavery in which Africans were forced to extract their raw materials so colonialists could fuel industrialization and capitalism in their home countries \cite{mhlambi2020from, ndjungu2020blood, day2023data, shilongo2023creativity, birhane2020algorithmic}.

Finally, the last vestige of colonial power to be challenged in African RDS are dependent partnerships. Africa currently lacks the technical infrastructure for large-scale data-driven technology, which pushes data scientists towards unfair agreements with powerful organizations to access vital resources \cite{shilongo2023creativity, hountondji2004producing, osaghae2004rescuing}. Even worse, companies such as Amazon, Google, Meta, and Uber use savior language such as ``liberating the bottom million'' to describe their digital services in Africa \cite{abebe2021narratives}. 

\textbf{Challenge Internal Power Asymmetry.}
\label{sec:chall_in}
Critical African philosophers conceptualize authoritarianism as governing to accumulate wealth and power rather than serving the needs of citizens and Africa as a whole \cite{nyerere1962ujamaa}.  
Even after liberation from colonial rule, many African philosophers accuse their governments of replacing the colonial ruling class instead of dismantling it \cite{coetzee2004laterMarx, kohnert2022machine}. To maintain their position, government officials focus on maintaining dependent relationships with the West and enforcing cultural nationalism to suppress dissent \cite{gwagwa2019recommendations}. 

African governments have already harnessed their control of national technology through internet shutdowns \cite{okolo2023responsible}. Therefore, to many authoritarian actors, powerful data technology is just another tool for suppression. Of particular concern to many African practitioners is China as a neocolonial collaborator with African authoritarian leaders and governments. Chinese companies have been found to provide the data technology Ethiopia, Uganda, and Zimbabwe have used to surveil their citizens \cite{okolo2023responsible}.

While authoritarian uses of data technology are resolutely unethical, more widely accepted uses of government data-driven technologies are scrutinized as well. 
The ubiquitous deployment of digital identification systems force citizens to choose between accessing important services or preserving their privacy from a system they have no control over \cite{gwagwa2019recommendations}. 
As governments consider adopting data technology, they need to be held accountable to their citizens \cite{ade-ibijola2023artificial,osaghae2004rescuing}. 
To combat the misuse of government power, authors in our corpus suggest data science initiatives should focus on improving government efficiency, transparency, and enforcement of citizens' freedom \cite{gwagwa2019recommendations, mabe2007security, eke2023towards}. 

\subsection{Assert Data Self-Determination}
\label{sec:Data Self-Determination}
Responsible African data science should be an avenue for bolstering the self-determination of Indigenous African communities.

\textbf{``For Africans, By Africans''.}
\label{sec:fubu}
This principle is inspired by the concerted efforts of African data scientists to reclaim leadership in African data science work \cite{chan2021limits}.
To combat deficit-based narratives about Africa, scholars in our corpus call on African data scientists to reclaim and celebrate their strength, rich cultures, and scientific achievements in conducting RDS \cite{abebe2021narratives,adelani2022masakhaner,hountondji2004producing, coetzee2004particularity, lauer2017african,gwagwa2019recommendations, carman2023applying}. 
Given the thousands of cultures that comprise Africa, African data scientists believe that grounding African data technology development in local talent, knowledge and data ethics can address local problems effectively and at scale \cite{eke2022forgotten,goffi2023teaching,coetzee2004laterMarx, shilongo2023creativity,day2023data,kohnert2022machine,coetzee2004african,segun2021critically, african_union2024continental, ruttkampbloem2023epistemic, gwagwa2022role, dignum2023responsible, olojede2023towards}.  

To achieve the ideal of African-led data science, the authors in our corpus highlight several changes to how African data scientists approach their work. African data should not be primarily collected for technology powers or published for immediate and uncontrolled use \cite{birhane2020algorithmic,hountondji2004producing}. Additionally, African data practitioners do not need tech superpowers to speak for Africans on the global stage, only provide off-the-shelf models, or oversee the standards of African data science institutions; Africans are more than capable of leading without interference \cite{ndjungu2020blood,abebe2021narratives,goffi2023teaching,mhlambi2023decolonizing, okolo2023responsible, ade-ibijola2023artificial, biko2004black}. This does not mean Africans should not collaborate with external data practitioners and vice versa \cite{hountondji2004producing, eke2023introducing}. 
Rather, local African data practitioners must lead data science work so its development is properly situated in the communities in which the technology will be deployed \cite{nwankwo2019africa,lauer2017african,kiemde2022towards, eke2023towards}.

\textbf{Treasure Indigenous Knowledge.}
\label{sec:treasure_ik}
With the legacy of colonial epistemic injustice, African modernization and Indigenous knowledge preservation are often viewed as at odds with each other \cite{african_union2024continental,kohnert2022machine,eke2023introducing}. On the contrary, many documents hold Indigenous knowledge as a pivotal component of responsible data science in Africa. As elders, griots, and other stewards of Indigenous knowledge pass,
younger generations are obligated to preserve their community's culture
\cite{kotut2024griot,ramose2004struggle}. 

Data-driven technology can be used to store Indigenous languages, customs, and history in close consultation with Indigenous communities. 
There are over 1500 languages indigenous to Africa, but very few are represented in data technology, such as natural language processing (NLP), which leaves out large portions of Africans from using ubiquitous technology \cite{shilongo2023creativity}. Local communities can never fully be represented if there is not an understanding of their roots or history \cite{ramose2004struggle}. Building datasets that represent Indigenous languages for inclusive models opens a whole set of new users who can digitally store and analyze Indigenous knowledge that is typically shared orally for future generations \cite{shilongo2023creativity,moahi2007globalization}. However, some scholars caution that releasing Indigenous information into the globalized data ecosystem could lead to a loss of cultural control and appropriation \cite{african_union2024continental, abebe2021narratives,eke2023towards, ade-ibijola2023artificial}. The boundaries of what Indigenous knowledge should be a part of data-driven technology must be understood by consulting with the community before proceeding on any project \cite{kotut2024griot,moahi2007globalization}.

African philosophers also emphasize Indigenous knowledge isn’t limited to the past \cite{hountondji2004producing}. Pre-colonial Indigenous knowledge needs to be reclaimed to develop African data values that reflect local communities \cite{abdul2023transhumanism, chan2021limits}. Investing in African responsible data science is an investment in creating new Indigenous knowledge \cite{uzomah2023african,lauer2017african}. Local talent do not have to reinvent the wheel to explore open questions in the more recent field of data science \cite{lauer2017african,mabe2007security, moahi2007globalization}. The richness of African knowledge can develop new responsible data science practices and understandings \cite{abdul2023transhumanism, mhlambi2020from, biko2004black, coetzee2004laterMarx}. 

\textbf{Protect Data Agency.}
\label{sec:data_ sov} 
Given the legacies of extractive colonialism, ownership is viewed as the key to data self-determination in Africa \cite{gwagwa2019recommendations,shilongo2023creativity,kiemde2022towards}. African ownership in the data science process can be achieved by codifying intellectual property rights \cite{african_union2024continental}, enforcing data ownership \cite{shilongo2023creativity}, and exploring Indigenous conceptions of collective privacy \cite{nwankwo2019africa,goffi2023teaching,mabe2007security,langat2020how, moahi2007globalization}. 

Africans are often regarded as ``simply'' data subjects \cite{shilongo2023creativity}. However, the role of a data subject is materially essential to data science work (without data, nothing can be done). Performance progress narratives that require unbounded data consumption devalue data subjects as dehumanized sources of raw material \cite{gwagwa2019recommendations,birhane2020algorithmic,olojede2023towards, mhlambi2020from, ndjungu2020blood}. This devaluing encourages data collectors to share and use data without the knowledge, consent, or compensation of data subjects \cite{sinha2023principlesafrofeminist, nyerere1962ujamaa}. Many African data ethicists call for a correction of this narrative to recognize data subjects as the proper owners by shifting power and access control to data subjects \cite{day2023data,ruttkampbloem2023epistemic, abdul2023transhumanism}. Achieving this shift in ownership should be done by demanding data-sharing terms and not working with data collaborators who do not honor these terms \cite{biko2004black, okolo2023responsible}. 
African ownership of data, resources for data science, and technical contributions are a non-negotiable for responsible data science in Africa.

\subsection{Invest in Local Data Institutions \& Infrastructures}
Prioritizing infrastructure, investing in local talent, and establishing sound policy and governance frameworks are essential for sustained and independent RDS in Africa.

\textbf{Invest in Physical \& Organizational Infrastructures.}
\label{sec:tech_infra}
To implement data technology in Africa, practitioners call for investment in physical data science infrastructure, assessment of the current capacities of technical infrastructure, and development of responsible data management practices \cite{moahi2007globalization, ruttkampbloem2023epistemic}. Achieving this principle in Africa is a big feat when electricity and broadband access is not only sparse but one of the most expensive to access in the world \cite{okolo2023responsible, ade-ibijola2023artificial}.  
Nigeria, Mozambique, and Rwanda have recognized the need to invest in technical infrastructure and have partnered with external tech companies and international financial institutions to build their respective capacities to host data-driven systems \cite{okolo2023responsible}. 
There are also innovative ways to work with current technical infrastructure to lessen reliance on external investment \cite{mhlambi2020from}. 
Technical infrastructure development should also coincide with the development of responsible data management protocols so African data and data science work are not vulnerable to dispossession \cite{abebe2021narratives, gwagwa2019recommendations}. 

\textbf{Cultivate Governance Infrastructure.}
The authors in our corpus call for sustainable and measured governance infrastructure to guide the development of data science in Africa \cite{african_union2024continental, chan2021limits}. Policy measures and regulations are major priorities for African data science communities. African Union member states are slowly developing data protection regulations, but many documents stress the urgency for African data policy \cite{kiemde2022towards, plantinga2024responsible}.
Without clear policies and legal standards for RDS, African data scientists lack guidance in their practices, leaving African communities vulnerable to data exploitation from external and internal actors alike \cite{abebe2021narratives,cisse2018look, mandaza2004reconciliationzimbabwe}. 
Governance infrastructures include incremental regulations \cite{gwagwa2019recommendations}, monitoring bodies \cite{goffi2023teaching}, continental commitments \cite{african_union2024continental}, and algorithmic impact assessments \cite{sinha2023principlesafrofeminist}. 

\textbf{Support Formal \& Informal Data Science Collectives.}
The African population has low attainment of digital skills \cite{okolo2023responsible, ade-ibijola2023artificial}. As large foreign technology companies set root in Africa, policymakers stress the need for monumental efforts to train local talent \cite{african_union2024continental,shilongo2023creativity, mhlambi2020from, cisse2018look}. Providing technical skills early in education will help prepare a strong cohort of future data scientists \cite{sinha2023principlesafrofeminist,nwankwo2019africa}. An indispensable part of a comprehensive data science education is data ethics \cite{goffi2023teaching, kiemde2022towards, ramose2004struggle}. A United Nations Educational, Scientific and Cultural Organization (UNESCO) survey found that very few African countries feel equipped to contend with the ethical implications of AI \cite{kiemde2022towards}. Teaching data ethics in Africa should involve centering the lived experiences and cultures of the students \cite{goffi2023teaching,kiemde2022towards}. Students should be educated about the common harms of data science and also develop their ethical discernment to prepare them for the sociotechnical complexities of data science. 

However, only focusing on supporting formal data science education (in which wider recognition and acceptance is a common issue related to epistemic injustice \cite{eke2022forgotten,chan2021limits}) neglects a large portion of potential data collaborators \cite{hountondji2004producing}. 81\% of jobs in Africa are based in informal economies \cite{shilongo2023creativity}. There should also be investments in integrating AI curricula in informal organizations like the Data Values Project to reduce educational barriers \cite{shilongo2023creativity}. There should be efforts to connect Africans interested in using data science for entrepreneurship \cite{biko2004black,shilongo2023creativity} and accessible data science job training \cite{abebe2021narratives}. Capacity-building in Africa necessitates the support of formal and informal data science collectives \cite{okolo2023responsible, abebe2021narratives}. 

Crucially, scholars in our corpus recommend dismantling the boundary between formal and informal data organizations to exchange technical knowledge, coordinate work, and pool resources \cite{dieng2023speaking, kling2023role}. 
Both forms of collectives have vital affordances and need to rely on each other to flourish. One form of collective is not meant to replace the other \cite{osaghae2004rescuing}. If both of these collectives are not supported, African data scientists will have to seek support outside of their communities, which furthers the ``brain drain'' of highly skilled Africans to global superpowers \cite{okolo2023responsible}. Investing in collectives also builds a workforce for in-house development which reduces foreign dependence \cite{kiemde2022towards, carman2023applying, plantinga2024responsible}. 

\subsection{Utilize Communalist Practices}
The development and deployment of data-driven technology should mitigate harms, involve communities in decision-making, and ensure reciprocal benefits for African stakeholders.  

\textbf{Ground Design in Community-Engaged Consensus.}
Consensus-building is a well-practiced strategy from African communities that can inform responsible practices and encourage effective collaboration \cite{wiredu2004moralfoundations}. Rather than majority-rule common in European societies, African elders discuss issues until they all agree on a final decision \cite{wiredu2004akan,carman2023applying}. Achieving consensus requires the final decision to be 1) the dominant view of the group, 2) in line with the common good, and 3) aligned with the morals of the individual parties \cite{coetzee2004particularity}. Consensus should be broached in an environment of trust, practical reason, humility, openness, and respect for the viewpoints of all involved parties \cite{coetzee2004particularity,nwankwo2019africa,gwagwa2019recommendations,mhlambi2020from, okolo2023responsible, gwagwa2022role}. 

Community engagement provides spaces for consensus in the data science lifecycle to include more perspectives \cite{day2023data,mabe2007security}. Akan philosophies regard the community as an invaluable resource that guides how every individual lives \cite{wiredu2004akan,metz2021african, coetzee2004particularity, mhlambi2023decolonizing,gwagwa2022role}. Therefore, community input is crucial for constructing a full picture of technical requirements, especially in high-stakes domains \cite{sinha2023principlesafrofeminist,mhlambi2020from, eke2023towards}. The concept of community can be misappropriated to deem any collection of stakeholders as sufficient community representatives. African communitarian ethics define a community as individuals with a shared identity who are emotionally invested in each other \cite{nwankwo2019africa,sinha2023principlesafrofeminist, ruttkampbloem2023epistemic, gyekye2004person}. With this more narrow definition of community, involving affected communities in all stages of the lifecycle requires building trust and respecting boundaries by gaining an understanding of cultural norms \cite{abebe2021narratives, ade-ibijola2023artificial}. Additionally, community members should be sufficiently trained or educated on the nature of the technology so they can provide well-informed input \cite{shilongo2023creativity,adelani2022masakhaner, plantinga2024responsible}.

Consensus processes should also include procedures for documentation to keep track of disagreements, dissenting opinions, and the progression of project values \cite{kling2023role}. It is not easy to achieve these conditions, so conflict management, negotiation, and reasonable bargaining are helpful mechanisms to fully consider and resolve contradicting positions \cite{gwagwa2019recommendations, osaghae2004rescuing, dieng2023speaking}. Consensus should be a dynamic feedback loop to ensure every contributor is on the same page about the team's approach to RDS \cite{segun2021critically, nwankwo2019africa, uzomah2023african, kohnert2022machine, abebe2021narratives, gwagwa2019recommendations, dignum2023responsible}. Community-centered consensus-building is a co-creation process in which all stakeholders depend on each other \cite{langat2020how, kohnert2022machine, abebe2021narratives, adelani2022masakhaner,nwankwo2019africa, lauer2017african, kiemde2022towards}.

\textbf{Instantiate Reciprocal Relationships.}
In many African philosophies, reciprocity is the foundation of a healthy society. In Akan society, practicing reciprocity ensures that community needs are met, while building deep social bonds \cite{wiredu2004moralfoundations}. African perfectionist proponents go as far as to assert that assisting others in achieving their goals makes someone more of a person \cite{wareham2021artificial,wiredu2004moralfoundations}. Without reciprocity, society becomes imbalanced and co-dependent \cite{coetzee2004particularity,mhlambi2023decolonizing,nyerere1962ujamaa}. 
There are numerous examples of African data subjects not reaping any benefits from the data collaborations they participate in \cite{gwagwa2019recommendations,abebe2021narratives}. This often leads to technically mediated harms while the controllers of data amass profits \cite{gwagwa2019recommendations}. 
Therefore, sustainable RDS should practice reciprocity on several dimensions \cite{wiredu2004moralfoundations}. If someone contributes to DDT they should meaningfully benefit from the system or project \cite{mhlambi2020from, gyekye2004person, sinha2023principlesafrofeminist}. Inspired by philosophies such as Ubuntu or Ujamaa, DDT should operate in a manner that benefits the society in which they are created and deployed \cite{eke2023towards, adelani2022masakhaner,dignum2023responsible}. 

Given the current gap between Africa's AI readiness and growing interest in AI adoption, many concede external partnership as a necessity~\cite{eke2023introducing, african_union2024continental}. However, exploitative external relationships set a precedent that curtails African self-determination in data science work~\cite{ndjungu2020blood,sinha2023principlesafrofeminist}. When building relationships, there are established obligations that each collaborator owes to the other~\cite{coetzee2004particularity, metz2021african}. Data collaborations must be predicated on trust, fair attribution of work, and a commitment to prioritizing the agency of African collaborators \cite{adelani2022masakhaner,nwankwo2019africa,abebe2021narratives,gwagwa2019recommendations, wareham2021artificial, wiredu2004moralfoundations}. 

\textbf{Mend Data-Driven Harms.}
When disagreements, conflict, or harm occur at any stage of the data science lifecycle, African data ethicists assert the need for mechanisms of accountability and reconciliation to correct wrongs and empower those impacted. In African societies, harm is not just actively making someone's life worse but also neglecting obligations to the community \cite{wiredu2004akan, gyekye2004person}. A person who causes harm is viewed as a moral failure who must be corrected by their community through sanctions and even mental rehabilitation to address deeper issues connected to their poor actions \cite{wiredu2004moralfoundations, coetzee2004particularity}. Even the most powerful members of society, such as chiefs, are subject to correction and even dismissal by their community for misconduct \cite{wiredu2004akan}. 

The adoption of AI and other DDT have already caused harm to African populations by way of data bias, socio-economic risk, and privacy violations \cite{ade-ibijola2023artificial}. 
There are African data ethicists who stress the need to develop procedures for communities and individuals harmed by DDT to seek restitution \cite{gwagwa2019recommendations,mhlambi2023decolonizing}. These solutions are dependent on African governments and external multinational organizations committing to transparency, equality, and restorative practices \cite{gwagwa2019recommendations,african_union2024continental,okolo2023responsible, dignum2023responsible, kiemde2022towards}. African governments can mitigate data harm by being transparent about their potential data collaborations, outlining their plans for data protection before, during, and after the deployment of DDT, and enforcing mechanisms of accountability and dissent from their citizens \cite{shilongo2023creativity, sinha2023principlesafrofeminist}. Similar to the dismissal of chiefs, powerful stakeholders acting outside of their agreed duties, must experience restorative consequences, not just a slap on the wrist \cite{mandaza2004reconciliationzimbabwe, ndjungu2020blood, shilongo2023creativity, mhlambi2023decolonizing, langat2020how, gwagwa2019recommendations, mhlambi2020from, biko2004black, coetzee2004laterMarx}.

\subsection{Center Communities on the Margins}
Community involvement ensures African DDT considers the needs and potential impacts of rural communities, women, youth and populations beyond the end-users. 

\textbf{Design in Solidarity.}
Solidarity is understood as looking out for other diverse communities based on mutual respect and the goal of social cohesion \cite{gwagwa2019recommendations, mhlambi2020from}. In Ubuntu understanding, solidarity is a deep care for others, including people of the past, present, future, and the environment \cite{mhlambi2023decolonizing,gwagwa2019recommendations, okolo2023responsible, dignum2023responsible, gwagwa2022role}. With this perspective, African data scientists in our corpus call for DDT to not be developed with only the end user in mind but all other communities that could be impacted by the technology
\cite{gwagwa2022role,olojede2023towards,gyekye2004person}.

While Africa needs to be included in global data science efforts, Africa itself is home to vastly diverse communities that should also be appropriately represented in these efforts \cite{adelani2022masakhaner,gwagwa2019recommendations,goffi2023teaching}. African communities' underrepresentation in datasets across all data science tasks is due to, as Gwagwa described, being ``uncounted, unaccounted, and discounted'' \cite{gwagwa2019recommendations}. Leaving communities out of data also excludes them from the benefits DDT provide \cite{gwagwa2022role}. Given the need to build explicitly African DDT, the lack of African datasets is a threat to efficacy \cite{okolo2023responsible,olojede2023towards, ade-ibijola2023artificial}.
Including marginalized communities requires mutual respect for diverse perspectives and creating procedures such as impact assessments to provide opportunities for inclusive input \cite{african_union2024continental, abebe2021narratives, goffi2023teaching, mhlambi2020from}. In addition, it’s important to challenge the social, political, and economic dynamics that push communities to the margins in the first place \cite{kiemde2022towards, olojede2023towards,segun2021critically,day2023data,uzomah2023african,abebe2021narratives}.

Solidarity violations between African countries is of particular concern. The success of one African community should not be predicated on the suffering of another \cite{ndjungu2020blood, biko2004black}. Upholding solidarity means that all actions made in the data science lifecycle should explicitly protect or improve the lives of vulnerable or marginalized communities. Exploiting the vulnerability of another is not only unethical but unsustainable due to our interconnected nature. The suffering of one community will eventually lead to the destruction of all communities \cite{nwankwo2019africa}.  
Banding together, ``watching one another's back,'' and developing DDT as a united front is key to mitigating harm \cite{olojede2023towards,nyerere1962ujamaa}.

\textbf{Center Remote \& Rural Communities.}
Development, especially technical development, is usually focused in urban centers and excludes remote and rural communities \cite{ade-ibijola2023artificial, sinha2023principlesafrofeminist, osaghae2004rescuing}. Given the lack of infrastructure in remote and rural communities, DDT should be used to develop and optimize infrastructures and public services for these regions \cite{carman2023applying, african_union2024continental}. 
However, it’s important to keep in mind that RDS done on behalf of rural and remote communities that do not consider their culture, livelihoods, and direct input can lead to harm \cite{african_union2024continental, ndjungu2020blood, carman2023applying}.

\textbf{Center Women.}
\label{sec:center_women}
Due to the prevalence of patriarchy in many African societies, authors in our corpus encourage the agency of women in data science efforts. A few documents suggest that women-led technology businesses and the education of women and girls should be incentivized \cite{african_union2024continental}. However, open questions remain about how to maintain African women's participation in a field known to be male-dominated and antagonistic to women \cite{african_union2024continental, gwagwa2019recommendations}. 

Afro-feminists have a response to the techno-chauvinism that dominates data science \cite{sinha2023principlesafrofeminist}. Rather than centering women in general, there must be a recognition of the intersectional status of African women \cite{sinha2023principlesafrofeminist}. As articulated by Rosebell Kagumire, African women experience domination through systems of patriarchy, race, sexuality, and global imperialism \cite{dieng2023speaking, coetzee2004particularity}. Therefore, data-driven technology should be developed with the complex needs of African women in mind, because their compounded experiences of marginalization provide insight into the needs of various oppressed populations \cite{sinha2023principlesafrofeminist}. 
There are numerous examples of African women harnessing the internet to fill in the gaps of an oppressive society and DDT holds similar potential \cite{dieng2023speaking}.
For example, Chil AI Lab Group is a women-led data science collective successfully using data technology to address the often neglected health needs of women in Africa \cite{eke2023towards}. 

\textbf{Protect \& Empower Youth.}
Africa is a young continent with a large population of educated and digitally native youth \cite{nwankwo2019africa, goffi2023teaching}. 
Prioritizing the youth of Africa is a two-pronged principle: 1) protect young people from harm and 2) empower youth to lead data science agendas. 
The youngest generation has a tech-savviness transferable to data science skills \cite{african_union2024continental,birhane2020algorithmic,abebe2021narratives}. 
If youth are expected to be the first adopters of African data technology then these systems should be designed to protect youth so they cannot be taken advantage of. 
Their comfort with technology may lead them to uncritically adopt a ``move fast and break things'' approach \cite{abebe2021narratives, ruttkampbloem2023epistemic}. To address these concerns, data science work should be intergenerational. African data technology should enrich the development of African youth and empower them to innovate, imagine, and contribute to bettering the communities they are a part of. 

\subsection{Uphold Common Good}
Ethical development and deployment of DDT requires a commitment to upholding fundamental human dignity and ensuring these technologies benefit all. 

\textbf{Adopt Technology with a Measured Mindset.}
Various scholars in our corpus want the development of African data science ecosystems to be balanced, measured, inclusive, and community-minded \cite{kohnert2022machine, eke2023towards, ade-ibijola2023artificial}. Without this approach, the adoption of AI and other data technologies can lead to more unrest and inequality across Africa.
To many, the potential of DDT is profound and would change the trajectory of African development \cite{african_union2024continental, mabe2007security}. Data are viewed as the driving resource for the Fourth Industrial Revolution \cite{carman2023applying, gwagwa2019recommendations}. There are African data scientists and governments who insist joining the AI boom will provide Africa the quality of life benefits afforded to the major players of past industrial revolutions \cite{okolo2023responsible, kohnert2022machine, coetzee2004laterMarx}. 

However, there is skepticism towards wholeheartedly diving into large-scale data science adoption \cite{uzomah2023african, olojede2023towards}. 
There is a need to quell the AI hype as the solution for all African problems and consider who will actually be served: Africans or the external powers propelling the AI boom \cite{wareham2021artificial,birhane2020algorithmic, sinha2023principlesafrofeminist}. 
Through the paradigm of measured development, technical development should move at the pace of social development \cite{kiemde2022towards, nyerere1962ujamaa}. Paulin Hountondji's critique of science in Africa applies well to data science development. Development should not be driven by ``scientific extroversion'' or catching up with global superpowers \cite{hountondji2004producing,goffi2023teaching}. Rather, the development of data science should be an investment in the progress of African people based on African intellect, priorities, and visions of the future \cite{shilongo2023creativity,biko2004black}.

\textbf{Preserve the Dignity of Data Contributors.}
Every human and community deserves humane treatment, and African scholars in our corpus do not want data technology to ever violate human dignity \cite{olojede2023towards,mhlambi2020from}. The African Charter on Human and Peoples’ Rights and the Universal Declaration on Human Rights set the precedent for the just treatment of humans \cite{african_union2024continental}. Regardless of these laws, African philosophies necessitate respect for human dignity because humans should be inherently valued for their existence and connection to others \cite{segun2021critically,metz2021african, dignum2023responsible}. Every human must be treated with respect, care, and concern for their well-being \cite{wiredu2004akan,dieng2023speaking,coetzee2004particularity, gyekye2004person, wiredu2004moralfoundations}. 
In applying the principle of universal dignity to RDS practices, every person involved in the data science lifecycle should be respected.
Individuals should not be used as a means to execute data work \cite{metz2021african,ramose2004struggle}. Rather, all efforts should be taken to ensure their well-being and dignity are 
preserved when asked to contribute to DDT \cite{gwagwa2019recommendations,abebe2021narratives}. This same respect also extends to communities. Collective agreements need to be honored, and collective work or resources should not be used in a manner that threatens the well-being of the community \cite{moahi2007globalization}. While this principle is self-evident, there are many cases in which the rights of Africans were violated for large-scale DDT \cite{african_union2024continental, segun2021critically, kohnert2022machine, moahi2007globalization}.

\textbf{Strive for Common Good with Systemic Change.}
African scholars view responsible data science as contributing to the safety, health, and goodness of all \cite{olojede2023towards}.
In various African philosophies, a person is defined by their commitment to acting for the benefit of those around them \cite{african_union2024continental, nwankwo2019africa, mhlambi2023decolonizing,coetzee2004particularity,ruttkampbloem2023epistemic, gyekye2004person,abdul2023transhumanism, wiredu2004moralfoundations}. 
DDT have to be made with the explicit goal of improving society and dismantling systemic harms \cite{sinha2023principlesafrofeminist, okolo2023responsible, eke2023towards}.  
In Africa, improving the efficacy of agriculture practices, healthcare access, responsiveness of public services, and the security of financial services are over-arching priorities \cite{carman2023applying,kohnert2022machine}. Achieving common good involves incorporating collective values early in the process \cite{langat2020how,dignum2023responsible}, guiding development with regulatory toolkits \cite{olojede2023towards}, not focusing on individualistic profit maximization \cite{gwagwa2019recommendations,segun2021critically,mabe2007security,nyerere1962ujamaa, mhlambi2020from,dieng2023speaking}, and encouraging the open sharing of data \cite{gwagwa2019recommendations,abebe2021narratives,day2023data}. The ultimate goal of RDS in Africa is upholding common good \cite{carman2023applying, metz2021african, olojede2023towards, mabe2007security}. 

\textbf{Maintain Harmony with Stakeholders.}
Scholars in our corpus urge for DDT to further the mutual well-being of all stakeholders. In addition, data standards and frameworks will be most effective when they harmonize with each other \cite{gwagwa2019recommendations, kiemde2022towards, wareham2021artificial, mabe2007security}.
In many African philosophies, harmony is not a state but a dynamic and reciprocal process of calibrating one's actions in response to changes in the environment. In Ubuntu ethics, dogmatism is rejected because it impedes individuals from acting in harmony with the changing world \cite{ramose2004ethicsofubuntu}. In Akan philosophy, morality is defined as acting in line with collective human interests \cite{wiredu2004moralfoundations}. 
Upholding harmony in data science can be understood through two dimensions: impact and practice. 

Data should be harnessed to bring people closer to their environment so they can act in the best interests of not only themselves but also those around them. In terms of practice, data ethics frameworks are most effective when all the elements of data science work are accounted for \cite{kiemde2022towards,gwagwa2019recommendations}. Also, acknowledging the unique ethical needs at each stage of the data science lifecycle can inform an adaptable practice of RDS. As Gwagwa, et al. assert, the harmonious practice of RDS in Africa requires country-level data ethics frameworks to be in alignment with frameworks developed at the continental level \cite{gwagwa2022role}. 

%% file: sections/05_discussion.tex
\section{Orienting African Data Ethics}
Now, we orient the African data ethics framework we proposed in the previous section by placing it in conversation with larger data ethics and African philosophy discourses. First, we summarize moral imperatives raised in our framework by connecting political analyses of African ``development''. Next, we code data ethics framework papers from different data science communities to explore how our framework relates to the larger data ethics discourse. Finally, we highlight how our framework raises a new perspective on what it means to operationalize data ethics.

\subsection{The Role of ``Development'' in African Data Ethics}
Rethinking power relations is the root of all of the values presented in our framework. Each principle offers two-fold analyses of how power relations in African data science hamper responsible practices. On one hand, neocolonial forces bring harm to Africans in the data science lifecycle and assert control that robs Africans of agency over their DDT. On the other hand, there are internal power asymmetries in existing sociopolitical dynamics that push fellow Africans to the periphery and can lead to further marginalization.  
By confronting these two dimensions of power asymmetries, African data ethics shows how RDS is not only an approach to technical work but an avenue for progressive power redistribution. 

The focus on self-organizing power seamlessly connects to classic African scholarship about development in Africa. Scholars such as Walter Rodney in \textit{How Europe Underdeveloped Africa} presents the European extraction of African resources, exploitation of African people and suppression of African state-building as a means to keep Africa in a state of limited development \cite{rodney1982europe, ndlovu-gatsheni2015genealogies}. However, other scholars acknowledge the impact of colonialism and imperialism but place the responsibility for Africa's lack of development on African leaders \cite{gumede2015exploring}. The elitism, corruption, and power-hungry tendencies of African leaders harm everyday citizens and the sustainable progress of their nations \cite{ahmed2014chapter}. Rather than choosing a side in this debate, our framework discourse demonstrates both problems must be overcome concurrently for progress to be made. 

Many scholars go a step further to even problematize the concept of ``development'' because the metrics for being ``developing'' or ``developed'' are set by global superpowers \cite{murrey2023beyond}. Therefore, to imagine responsible African data science beyond AI readiness metrics, it's important for African scholars to define what success looks like to them. Our analysis of the African data ethics corpus surfaces a set of clear goals for future African data science grounded in moral clarity and ideological traditions. These value-laden objectives guide and continue to guide the critiques, initiatives, and policies Africa develops for RDS. For examples of African data science initiatives aligned with the principles of our framework, see Appendix ~\ref{apdx:more_case_studies}.

\subsection{Comparing Across Particularist Data Ethics Frameworks}
\label{sec:compare}
We compare our framework to seven particularist frameworks to gauge the breadth of our work alongside: intersectional feminists \cite{klein2024data}, Western technology powers \cite{floridi2018ai4people}, Indigenous communities of Turtle Island \cite{carroll2021operationalizing}, policymakers from global superpowers \cite{jobin2019global}, Central Asian data scientists \cite{younas2024proposing}, Muslim data scientists \cite{raquib2022islamic} and Black American data scientists \cite{monroe-white2021emancipatory}. We call these works particularist frameworks because the authors of each paper aim to present the data science values of distinct global communities and apply the ethical theories of distinct philosophical traditions. The first author reviewed each framework and used the minor principles in our African data ethics framework as a codebook (see Appendix ~\ref{apdx:compare_table} for comparative coding results).

Each framework covered, at most, 68\% of our proposed 19 principles. Delving into the diverging and converging principles provides insight into how African data ethics deepens global data ethics discourse through particularism. 

None of the frameworks discussed \textbf{Centering Remote and Rural communities}. Our framework highlights RDS for rural communities because they represent a significant portion of Africa and have unique needs not fully met by the status quo \cite{barrett2017structural}. Communities have distinct shared experiences that inform their values and normative understandings of the world. Engaging with data ethics from different cultural standpoints exposes data scientists to approaches or potential harms they would have never considered \cite{adamu2021rethinking}.

\textbf{Strive for Common Good with Systemic Change} and \textbf{Ground Design in Community-Engaged Consensus} were the only principles discussed by all the frameworks. This finding also falls in line with a popular understanding of philosophy: universalism. The whole RDS community is broadly guided by a universalist commitment to social good \cite{floridi2018ai4people}. However, upon closer review, references to the common good in non-African frameworks often remain abstract, lacking the depth and specificity seen in the work of African data ethicists. African appeals to universalism are grounded in the tradition of speaking truth to power.
Drawing from critiques of international human rights law, African scholars highlight the importance of pairing narrative restoration—the recognition and reclamation of African humanity and agency—with material restoration, including tangible reparative actions \cite{biko2004black, oyowe2014african, gordon2022universalism}. Our framework corpus notes numerous examples of data science practices in Africa violating the knowledge, ownership, and safety of African communities. Social good is often taken for granted as an obvious universal goal for RDS. However, African data ethics urges the RDS community to confront the harm they have caused before it can truly achieve social good. 

\subsection{Reframing the Operationalization of Data Ethics} 
The primary critique of data ethics frameworks is the difficulty of operationalizing ideals into effectual protocols or technical practices \cite{munn2023uselessness}. We offer a new perspective on the utility of data ethics frameworks: constructing frameworks is an essential operation in itself for RDS.

Articulating values and moral convictions are the first steps to contextualizing responsible data science in the epistemic traditions implicitly embedded in current practices, the socio-political status quo, and the future worlds to build for. Surfacing these contexts is important because they afford data scientists the ability to make technical decisions, develop projects, and outline long-term goals with a clear vision. The African scholars in our corpus recognize that if they don't set the moral vision for data science in their communities, African data science will implement the divergent and often times destructive vision of global superpowers. For example, Western longtermism imagines a world in which DDT dominates humanity \cite{gebru2024tescreal}, while Afrofuturism imagines a future in which DDT joins with humanity to build a more just world \cite{abdul2023transhumanism}. Articulating optimistic, collectivist, and resolutely African moral visions such as Afrofuturism is crucial for Africa as an emerging site of data science. Thus, as responsible data scientists, we must take responsibility for envisioning and implementing futures for ourselves and our communities. Otherwise, we remain at risk of perpetuating the injustices we seek to mitigate with RDS.

%% file: sections/06_conclusion.tex
\section{Limitations}
Our reflexive thematic analysis should not be conflated with a systematic literature review. While we cast a wide net to build our corpus (including gray literature), there are impactful African data ethics documents not included in our framework. Therefore, the resulting framework should not be considered comprehensive but an introduction to notable ideas. Future work will aim to incorporate more African data ethics documents and engage African practitioners to evaluate the merits, gaps, and usability of our framework. 

\section{Conclusion}
Through a thematic analysis of 50 documents, we derived an African data ethics framework that encompasses six major principles: 1) Challenge Power Asymmetries, 2) Assert Data Self-Determination, 3) Invest in Local Data Institutions \& Infrastructures, 4) Utilize Communalist Practices, 5) Center Communities on the Margins, and 6) Uphold Common Good. Our framework scratches the surface of African data ethics discourse, and the surface is rich with historically grounded, communitarian, and pragmatic insights for RDS.

A comparative analysis of our framework with seven other data ethics frameworks highlights African perspectives as progressive and needed voices in global data ethics discourse. For truly pluralistic and responsible data science, we urge the RDS community to readily seek the moral perspectives of Africans, other practitioners of the Global Majority, and the particular data science communities to which they belong. Such reflexivity will not only enrich the theoretical foundation of data ethics but can inform more equitable and culturally responsive approaches to data governance, algorithmic fairness, and technological development. 

%% file: sections/07_appendix.tex
\section{PRISMA Diagram of Document Collection}
To supplement the description of the data collection process in \nameref{sec:methods} section, the following PRISMA diagram illustrates the corpus-building process including the identification, screening and inclusion stages. 
\label{apdx:method}
\begin{figure}[!ht]
  \centering
  \label{fig:prisma}
  \includegraphics[width=\linewidth]{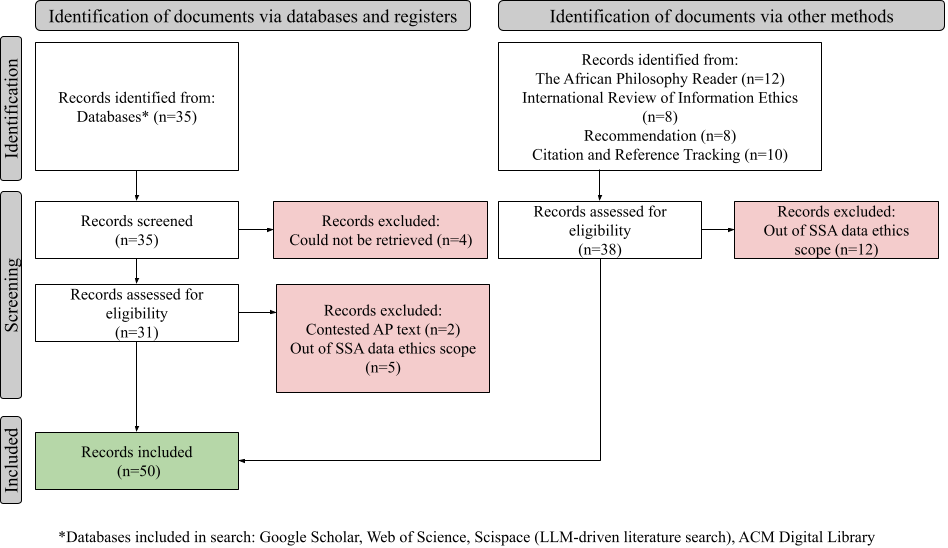}
  \caption{PRISMA diagram of document collection process for African data ethics corpus.}
  \Description{There is a note for the ``identification of documents via databases and registers'' that states: Databases included in search: Google Scholar, Web of Science, Scispace (LLM-driven literature search), ACM Digital Library and IEEExplore. The PRISMA flowchart starts with the identification of documents from databases and registers or other methods and documents of pruned due to exclusion criteria described in the methods data collection section until a final count of 47 documents are reached.}
\end{figure}

\section{African Data Ethics Framework Case Studies}
\label{apdx:more_case_studies}
The following section presents six case studies from the African data ethics corpus to further contextualize our framework with real world examples from data science work in the African context.

\subsection{Challenge Power Asymmetries Case Study: DRC Mineral Extraction}
The violent exploitation of miners in the Democratic Republic of Congo (DRC) is a harrowing example of how unchecked power from the West and within Africa corrupts the data science ecosystem. The DRC is home to an abundance of minerals necessary for data science. Specifically, cobalt and silicon are foundational components of all technology, especially the vast number of computers that store and process data in global data centers \cite{ndjungu2020blood}. To keep up with computing demand, multinational companies collect a copious amount of minerals from the DRC with feigned ignorance of the unsafe working conditions, the prevalence of child labor, and the inhumane actions of the rebel groups who often control mining operations. Silicon and cobalt are often referred to as ``blood minerals'' because Western companies are able to make billions of dollars from the technology industry while the DRC continues to experience violent internal displacement, ineffective interventions, and a minuscule fraction of the value of their mining labor \cite{ndjungu2020blood}. The extractive relationship between the Congolese and large Western companies has direct parallels to DRC's colonial relationship with Belgium \cite{ndjungu2020blood}. The monarch of Belgium, King Leopold, violently claimed DRC to extract and sell raw materials so Belgium could be a major player in meeting the material demands of an industrializing Europe without any concern for the humanity of the Congolese people. This colonial relationship is furthered by African leaders who have assumed the role of middlemen in the mineral trade. Dodd-Frank Section 1502 is a United States law passed to address dehumanizing mining labor practices \cite{ndjungu2020blood}. This law required companies to execute due diligence measures to ensure they were not selling DRC minerals mined from conflict. Rather than adhering to this law, multinational companies pulled out of direct agreements with the DRC and joined new partnerships with neighboring countries such as Rwanda. These countries then acted as middlemen, buying conflict minerals from the DRC and selling them to multinational companies, allowing the companies to legally distance themselves from the exploitative sourcing\cite{ndjungu2020blood}. Understanding the colonial and modern-day political background of DRC blood minerals is key to contextualizing calls for reducing the scale of datasets and demand for new technology \footnote{\url{https://newint.org/violence/2024/its-time-hold-big-tech-accountable-violence-drc}}. Every leader within the data science ecosystem has the responsibility to challenge and not perpetuate colonial and asymmetric power. 

\subsection{Assert Data Self-Determination Case Study: The Promise of African-led Data Science}
Masakhane is an African natural language processing (NLP) collective that builds language datasets and models in Indigenous African languages. By all reports across documents, Masakhane practices all the principles proposed in this framework and especially upholds the major principle of Data Self-Determination \cite{eke2023responsible, chan2021limits, shilongo2023creativity}.They uphold these principles with a commitment to centering African values as their founding principles, working with existing public datasets to not infringe on data privacy, and building language datasets to preserve indigenous African languages for the future to come \cite{adelani2022masakhaner}. Masakhane also has a very welcoming and communal organizational structure that includes any interested party in weekly meetings and communications. Once a member wants to contribute to a Masakhane project, they must undergo in-house training to maintain quality of the their dataset and model. They also explicitly prohibit ``parachute research from the Global North'' to ensure the time and resources of their collective provide direct benefit to their communities.\footnote{\url{https://www.masakhane.io/}} Finally, every project plan includes a discussion of data privacy considerations to guide their work. While there are many other parts to Masakhane's work, the practices described in their public documents demonstrate an African data science community that is closely aligned with the perspectives represented in our framework. Masakhane is part of the growing grassroots efforts to imagine and practice what African-led responsible data science can achieve.

\subsection{Invest in Local Data Institutions \& Infrastructures Case Study: Building the Capacity of National Statistical Offices}
As a formal data collective, the National Institute of Statistics of Rwanda (NISR) developed a report to guide their management procedures for administrative data \cite{habimana2018guidelines}. Their report recognizes the data sharing network they are a part of, raises concerns with data quality specific to Rwanda, and proposes new administrative standards for assessing data quality. While the authors recognize that their data management infrastructures need to progress, they view collaborations between local data practitioners as the key to development. Development collaborations include technical workshops, conferences, dissemination of data quality frameworks, and supporting staff in their respective data work \cite{habimana2018guidelines}. African data scientists are eager to develop their communities' capacity to manage data science projects. Organizations such as NISR recognize that this development requires a comprehensive assessment of the status quo, supportive collaboration, and incremental development of data standards.

\subsection{Utilize Communalist Practices Case Study: Challenging Utilitarian Data Ethics with a Communitarian Analysis}
Another aspect of practicing communalism in data science is applying communitarian theories as a lens for evaluating data ethics. The Western concept of utilitarianism is a predominant paradigm in data ethics. Utilitarian data science aims to construct AI and other DDT that maximize the amount of social good and minimize the amount of social harm at scale (see effective altruists). African communitarian theories provide novel and strong critiques of utilitarian data ethics as well. One African data ethicist, in particular, applies African values to outline why utilitarianism: 1) trivializes human dignity through rationality, 2) justifies the suppression of non-dominant people and values, 3) ignores the role relationality plays in human-AI interaction, and 4) misinterprets the nature of self-sacrifice \cite{metz2021african}. Respectful debate with diverging perspectives is essential to the progress of responsible data science.
 
\subsection{Center Communities on the Margins Case Study: Limits of Inclusive Representation in African Facial Recognition Technology}
African people, and Black people in general, are severely underrepresented in facial recognition datasets, and this has led to a performance bias against Black users. African technology companies committed to addressing this bias so their primarily African users could rely on their products. For example, a women-led facial recognition start-up called BACE curated a diverse dataset from the local community so their system could better detect Black subjects \cite{eke2023responsible}. Users upload photos of their IDs and short videos from their phones to confirm their identity \footnote{\url{https://www.bacegroup.com/}}. The technology was created to aid financial fraud investigation efforts in Ghana that are hampered by many citizens not having formal identification documents \footnote{\url{https://www.thehabarinetwork.com/meet-charlette-nguessan-she-and-her-team-have-innovated-facial-recognition-technology-designed-to-identify-black-africans}}. While at face value BACE is practicing community-engaged work, the principles described above call for more meaningful community involvement. Data representation is only one stage of the data science lifecycle that communities should be involved in. BACE was meeting the needs of financial institutions but did the team gauge if the local community was comfortable with this identification product \cite{birhane2020algorithmic}? To center all communities in responsible data science, local communities should be involved in the ideation, creation, deployment, and maintenance of data-driven technology to build trust and buy-in for new technology \cite{okolo2023responsible}. The work of N’Guessan is very aligned with the principles of data self-determination \& infrastructures because it was in indigenous effort to build a DDT for their local Black community. While this case study is seemingly doing everything ``right'' facial recognition is still used for surveillance and has performance issues for marginalized communities \cite{birhane2020algorithmic}.

\subsection{Uphold Common Good: International Partnership to Increase Access to COVID-19 Information}
At the beginning of the COVID-19 pandemic, the Rwandan government partnered with the German technology company GIZ to develop a chatbot for remote communities to access tailored COVID-19 information and guidance \cite{kohnert2022machine}. To meet the needs of Rwandan users, the chatbot can communicate in the local language of Kinyarwanda; the medical advice is based on the Rwandan medical databases, and the project is open source \footnote{\url{https://github.com/Digital-Umuganda/Mbaza-chatbot}}. Beyond the features of the product, both organizations worked together to develop Rwanda's technical infrastructure to not only host the chatbot software but also maintain local technologies in the future \footnote{\url{https://www.giz.de/en/workingwithgiz/KI-Ruanda-Digitalisierung.html}}. The RBC chatbot is the product of an equitable partnership of African and Western data organizations that were committed to promoting the well-being of their community in the face of a catastrophic pandemic that impacted the world. The collaboration was successful because their decisions were attuned to each stakeholder's capabilities and limitations \cite{kohnert2022machine}. By all accounts, this is an example of a harmonious, dignified, and socially good data science practice.

\section{Comparative Analysis Table}
As a supplement to the discussion's comparative analysis, the following table shows the coding of minor African data ethics principles to each of the seven particularist frameworks.
\label{apdx:compare_table}

\begin{figure*}
  \centering
  \includegraphics[width=\textwidth]{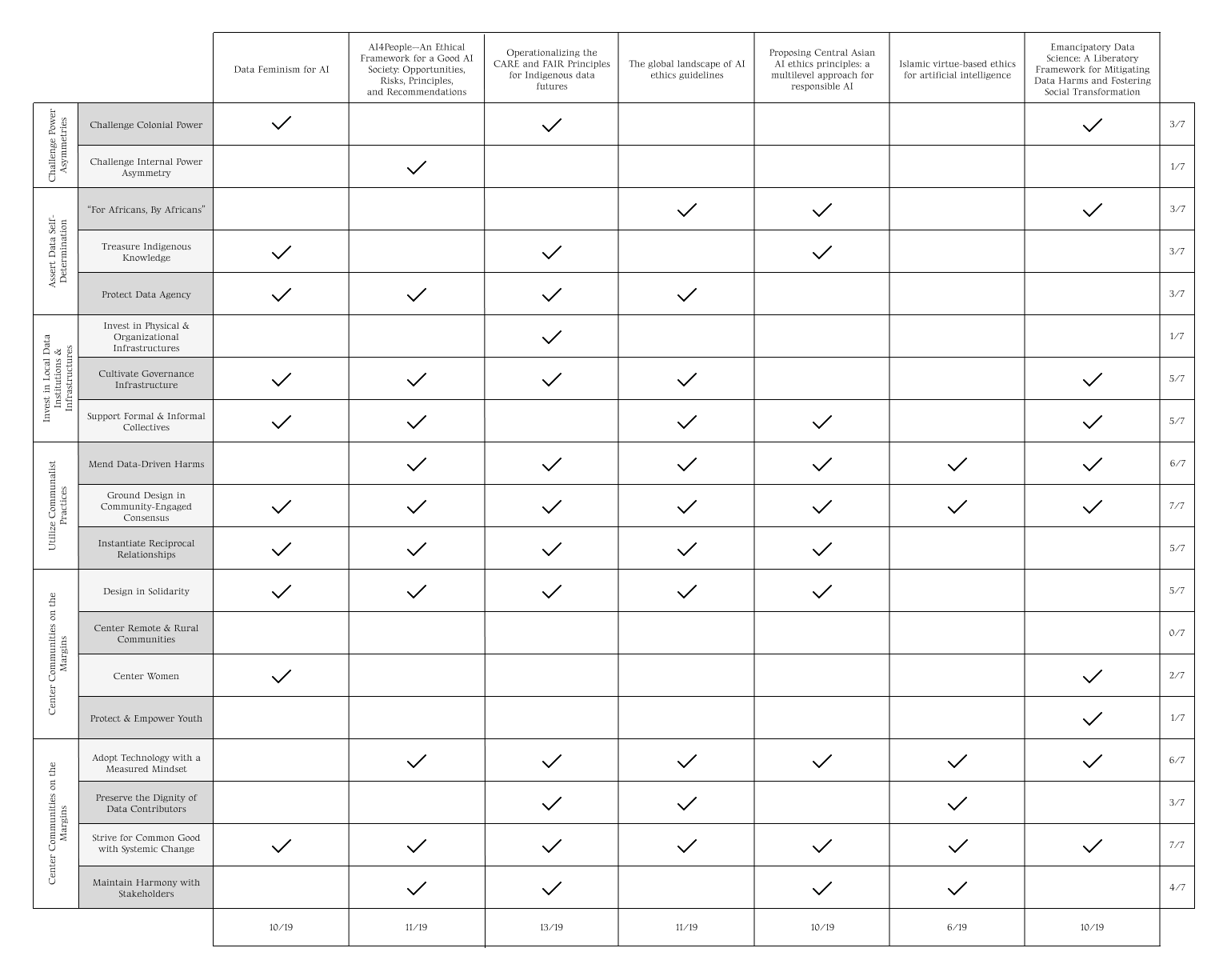}
  \caption{A table of minor African data ethics principles on the y-axis and seven comparison papers on the x-axis (\cite{klein2024data,floridi2018ai4people,carroll2021operationalizing,jobin2019global,younas2024proposing, raquib2022islamic, monroe-white2021emancipatory}. Between these axes a check mark indicates a principle is covered by the compared framework.} 
  \Description{At the last row there are fractions to count each checked cell by data ethics framework (Data Feminism for AI: 10/19, AI4People: 11/19, CARE \& FAIR Principles: 13/19, Global Landscape: 11/19, Central Asian AI ethics: 10/19, Islamic AI ethics: 6/19, Emancipatory Data Science: 10/19). In the last column, there are fractions to count each checked cell for each minor principle in an order consistent with the order of section 4 (3/7, 1/7, 3/7, 3/7, 4/7, 1/7, 5/7, 5/7, 6/7, 7/7, 5/7, 5/7, 0/7, 2/7, 1/7, 6/7, 3/7, 7/7, 4/7).}
\end{figure*}